\documentclass[twocolumn,showpacs,preprintnumbers,amsmath,amssymb]{revtex4}

\usepackage{graphicx}
\usepackage{dcolumn}
\usepackage{bm}
\usepackage{wrapfig}
\usepackage[latin1]{inputenc}

\begin{document}

\title{ Random deposition of particles of different sizes}

\author{F. L. Forgerini}
\email{fabricio_forgerini@ufam.edu.br}
\affiliation{ ISB, Universidade Federal do Amazonas, 69460-000 Coari-AM, Brazil }

\author{W. Figueiredo}
\email{wagner@fisica.ufsc.br}
\affiliation{Departamento de F\'isica, Universidade Federal de Santa Catarina, 88040-900 Florian\'opolis-SC, Brazil
}

\date{1 April 2009}

\begin{abstract}
We study the surface growth generated by the random deposition of particles of different sizes. A model is proposed where the particles are aggregated on an initially flat surface, giving rise to a rough interface and a porous bulk. By using Monte Carlo simulations, a surface has grown by adding particles of different sizes, as well as identical particles on the substrate in (1 + 1) dimensions. In the case of deposition of particles of different sizes, they are selected from a Poisson distribution, where the particle's sizes may vary by one order of magnitude. For the deposition of identical particles, only particles which are larger than one lattice parameter of the substrate are considered. We calculate the usual scaling exponents: the roughness, growth and dynamic exponents $\alpha, \, \beta \,$ and $z$, respectively, as well as, the porosity in the bulk, determining the porosity as a function of the particle size. The results of our simulations show that the roughness evolves in time following three different behaviors. The roughness in the initial times behaves as in the random deposition model. At intermediate times, the surface roughness grows slowly and finally, at long times, it enters into the saturation regime.
The bulk formed by depositing large particles reveals a porosity that increases very fast at the initial times, and also reaches a saturation value. Excepting the case where particles have the size of one lattice spacing, we always find that the surface roughness and porosity reach limiting values at long times. Surprisingly, we find that the scaling exponents are the same as those predicted by the Villain-Lai-Das Sarma equation.

\end{abstract}

\maketitle

\date{01 April 2009}

\section{Introduction}

The research on the geometric properties of growing surfaces is one of the most important in the field of the non-equilibrium statistical physics, not only because it is a challenge for the theoretical physicist to model these properties, but especially for the intrinsic experimental interest in shaping the surfaces with desired purposes \cite{barabasi, meakin}. The random deposition is the simplest known model, where particles are aggregated onto an initially flat substrate. Because lateral correlations among the deposited particles are completely neglected, the continuous and discrete atomistic versions of the model have exact solutions. If a surface relaxation mechanism is allowed to the deposited particles in the random deposition model, height-height correlations naturally appear. Although the corresponding discrete model did not present an exact solution, it can be described by the linear Edwards-Wilkinson equation (EW) \cite{EW82}, which is exactly soluble. A generalization of the (EW) equation was proposed by Kardar, Parisi and Zhang in 1986, including a non-linear term that accounts for the lateral growth of the interface. Although the growth process is a local phenomenon, the lateral growth is related to the spreading of the heigth fluctuations along the surface. This is characterized by a correlation length that increases with time, and it reaches a maximum value corresponding to the linear dimension of the substrate. A perpendicular correlation length is also defined that is related to the fluctuations in heigth along the growth direction. The non-linear KPZ equation \cite{KPZ86} is very useful to describe porous deposits, as those ones generated by the ballistic deposition of particles \cite{Vold}.

The morphology of the surface is described by the interface width $w(L,t)$, which characterizes the roughness of the interface, is defined by the rms deviation of the height $h$ around is average value $\bar{h}$, that is, 

\begin{equation}
w(L,t) = \sqrt{<(h-\bar{h})^{2}>},
\end{equation}
where $L$ is the linear dimension of the substrate, $t$ is the time elapsed after the growth start, the overbar indicates spatial average and the angular brackets mean configurational averages. It is well established that a large class of growth models follow the Family-Vicsek scaling relation \cite{FV}

\begin{equation}
w(L,t) \sim L^{\alpha}f(\frac{t}{L^{z}}),
\label{family_vicsek}
\end{equation}
where the scaling function $f(x)$ is a constant when $x$ is very large, and $f(x) \sim x^{\beta}$ when $x << 1$. The exponent $\alpha$ characterizes the interface width, $z$ is the dynamic exponent, while $\beta$ is the growth exponent. These exponents are not independent and are related by $\alpha = \beta z$. 

Most of the studies performed in the area of surface growth focus their attention in the determination of these exponents, and whether they fit to the Family-Vicsek ansatz. These calculations have been done analytically, by solving stochastic differential equations or employing mean field approximations, and through extensive use of Monte Carlo simulations \cite {das_sarma, albano, drossel, Ber, horowitz_arvia, Reis, barato_oliveira}. In this work we use Monte Carlo simulations to study the surface growth in $(1+1)$ dimensions due to the deposition of particles of different sizes. There are in literature some studies where two or more different deposition models are combined \cite{cerdeira_95, elnassar_cerdeira} or two species of particles are deposited \cite{caglioti, trojan, karmakar} in order to describe the time evolution of roughness in real systems. Here, we choose our particles to be deposited from a modified Poisson distribution, with average size equals to five lattice spacings and maximum size equals to nine. This type of distribution appears to be relevant in some ash particles deposition on the heat exchange surfaces \cite{Baxter, lourival}. We also consider in this work, the deposition of identical particles, which size is larger than one lattice spacing of the substrate. The deposition of particles proceeds as in the pure random deposition model, however, correlations between columns naturally appear due to the deposition of particles larger than one lattice spacing. The random deposition of particles of unit size generates a compact bulk and an infinitely large interface width. This means that for a linear substrate of size $L$, $w(L,t)$ becomes infinitely large as the deposition time $t$ goes to infinite. On the other hand, the deposition of particles larger than one unit produces a porous bulk. Then, we can determine the evolution of the porosity of the bulk simultaneously with the scaling properties of the surface. We say that the interface width satisfying Eq. (2) is self-affine, which means that rescaling part of the interface anisotropically we obtain an interface that is statistically indistinguishable from the whole one. In order to characterize the morphology of a rough interface, it is sufficient to know the value of the exponent $\alpha$ \cite{barabasi}. We show that the surface growth presents three different behaviors as a function of time. At the initial times, it behaves as in the usual random deposition of particles of unit size. Then, at intermediate times, when the lateral correlations between columns develop, it grows more slowly, and finally reaches a saturation regime at long times. Porosity attains the saturation regime faster than the interface width, and it presents large values even for the smallest particle of size two lattice spacings. Finite-size effects are not observed for this property, and for particles of size corresponding to four lattice units of the substrate, the porosity is already $85\%$ of the maximum possible porosity, that is, $P = 0.5$.

The paper is organized as follows. In Sec. II we present the model, the deposition rules and some details concerning the Monte Carlo simulations. In Sec. III, we present the results for the interface width as a function of particles' sizes. Sec. IV contains the results for the scaling exponents when the deposited particles are identical, and Sec. V is devoted to the study of bulk porosity. Finally, in Sec. VI, we present our main conclusions.

\section{Model for deposition of large particles }

We propose a model where the particles are dropped randomly over a finite linear substrate, which is divided into cells of unit size. All the particles also have unit height and the columns where they land are increased by one unit. Particles of different sizes are selected from a modified Poisson distribution, with sizes changing by one order of magnitude. The most important difference between this model and the simple random deposition, where only particles of unit size are deposited, is that the present model naturally allows for correlations among the columns, that is, it leads to a lateral growth of the interface. We studied the effect of particles's sizes on the scaling exponents, as well as on the global porosity. In this study, we performed Monte Carlo simulations in two quite different conditions: 1) adding particles of different sizes, and 2) adding identical particles. For the deposition of identical particles, only particles which are larger than one lattice spacing of the substrate are considered. In the case of particles of different sizes, they are selected from a modified Poisson distribution.

The Poisson distribution was modified in this work in order to take into account the discrete nature of the particles and a maximum size of the particle to be deposited during our Monte Carlo simulations. The probability of a particle with size equal to $n$ to be selected for a deposition trial is 

\begin{equation}
 \mathcal{P}(n,x) = \frac{x^n e^{-x}}{a n!},
\end{equation}
where $n$ ranges from $one$ to $nine$ lattice spacings, which is the maximum permitted size for deposition in the present model. As we fixed $x$ at the value $five$, the normalization factor $a$ is equal to \textit{0.96143}. Therefore, this is a type of a modified Poisson distribution with a finite set of possible results. For a true Poisson distribution, where all values of $n$ are possible, $x$ should be the average value of the distribution and the normalization factor $a=1$. This modified Poisson distribution is believed to be realistic to describe some ash deposits, where particles's sizes may vary by two or more orders of magnitude, and are well represented by a Poisson distribution \cite{Baxter}.

When we perform simulations with particles of different sizes, the particles to be deposited are flat, height one, and may change from (1 X 1) to (9 X 1) in units of the lattice parameter of the linear substrate, according to the modified Poisson distribution. In the case of deposition of identical particles, we choose just one species of particle, with a defined size. The simulations in this work are performed in (1+1) dimensions, the resulting deposit is two-dimensional, and we always use periodic boundary conditions along the linear substrate. 

To add a particle on the substrate, a cell on the lattice is randomly selected. At this selected position we place the midpoint of the particle, trying to extend horizontally, at both sides of the initial position, this site to the full size of the particle. In the case of particles with even size, as 2, 4, 6 and 8 lattice spacings, we choose as their midpoint, the position 1, 2, 3 and 4, respectively.

We allow the reflection of a particle by the surface and, in this case, the trial of deposition is lost, the particle is not incorporated to the substrate. The reflection is due to the geometric constraint properties of the surface. For each particle deposition attempt, we look at the number of vacant sites around a randomly selected position on the substrate. The number of vacant sites around this site determines the possible deposition or reflection of the selected particle. If the vacant region is smaller than the size of the selected particle, the particle will be rejected. On the other hand, if the vacant region is larger or equal to the size of the particle, there are no geometric restrictions and the incident particle permanently sticks to that region.

\begin{figure}
\resizebox{6cm}{!}{\includegraphics[scale=0.5]{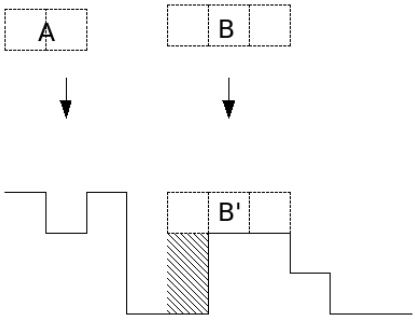}}
\caption{\textit{Illustration of the growth rules, where geometric restrictions lead to the reflection of a particle by the surface. Particle ``A'' will be rejected, while particle ``B'' will be added to the surface.}}
 \label{fig:figure1}
\end{figure}

The growth rules are ilustrated in Fig. \ref{fig:figure1}. The particle labeled by ``A'' will not be incorporated to the surface, while the particle labeled by ``B'' is added to the deposit. Under the ``B'' particle, a shadow zone is created, where no other particle can be aggregated. This is the mechanism that leads to the formation of a porous structure inside the bulk. In order to determine the global porosity, we calculate the quantity

\begin{equation}
 P = \frac{V_{pore}}{V_{pore} + V_{solid}},
\label{porosidade}
\end{equation}
where $V_{pore}$ is the volume occupied by the vacant sites, and $V_{solid}$ is the proper volume occupied by particles. The vacant sites here mean only the empty sites not belonging to the surface. The overhangs from the topmost layer are not considered vacant sites. In this work, we show that the deposition of particles larger than one lattice parameter of the substrate, generates a highly two-dimensional porous structure. The time evolution of the porosity and its dependence on the size of the deposited particles are presented in Sec. V.

\section{Particles of different sizes}

In this Section, we focus our simulations on the estimate of the scaling exponents and determination of the universality class of the model, when the deposited particles are selected from the probability distribution, Eq.(3). The usual scaling exponents: the growth, roughness and dynamic exponents, and also the global porosity in the bulk, are calculated for deposition of particles larger than one lattice spacing. Some earlier studies \cite{karmakar, Reis}, clearly point for the existence of two different regimes of growth, with growth exponents $\beta_1$ and $\beta_2$, when only two species of particles are deposited. The observation of the data for the time evolution of the roughness shows a change in its behavior, from an uncorrelated growth, whose exponent is $\beta_1$ to another one, where lateral correlations are present, which is characterized by a second growth exponent, $\beta_2$.

At the very initial times, the surface is free of lateral correlations, once the particles are incorporated to the substrate, following the rules of the random deposition model. The exponent $\beta_1$ is near $1/2$, which is the typical value for the random deposition model. However, at intermediate times, height-height correlations develop and we find a smaller value for the growth exponent $\beta_2$. Finally, at long times, the roughness enters into a saturation regime, and the surface is characterized by the roughness exponent $\alpha$.

\begin{figure}
\resizebox{6cm}{!}{\includegraphics[scale=0.45]{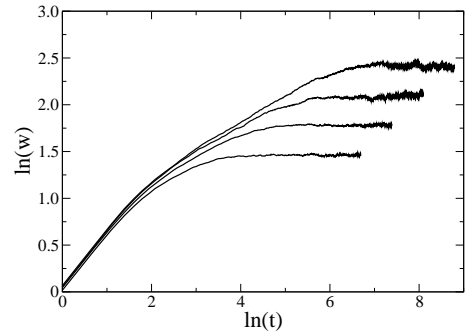}}
\caption{\textit{Log-log plot of the roughness versus time for deposition with particles' size taken from the modified Poisson distribution, Eq. (3). From bottom to top the lattice sizes are L = 1024, 2048, 4096, 8192.}}
 \label{fig:figure2}
\end{figure}

In Fig \ref{fig:figure2} we show the log-log plot for the time evolution of the roughness, where simulations were performed with particles selected from the Poisson distribution, Eq. (3), with $x=5$ and $n$ ranging from $1$ to $9$. The results we are showing, represent averages considering 10$^3$ different samples for the linear sizes L = 1024, 2048 and 400 samples for L = 4096 and 8192. In this log-log plot we see two different linear regimes at the initial times, where we find the value $\beta_1$ = 0.503 $\pm$ 0.005 for all the lattice sizes, which is close to the expected value for the random deposition model, whose exact value is $1/2$ in any spatial dimension.

\begin{figure}
\resizebox{7cm}{!}{\includegraphics[scale=0.4]{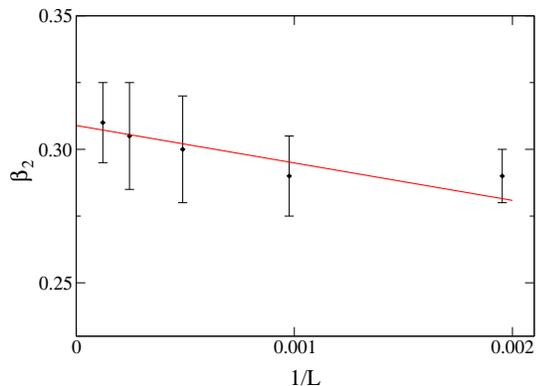}}
\caption{\textit{Plot of the growth exponent $\beta_2$ versus 1/L. The straight line is the best fit of the $\beta_2$, for several values of L and surface formed by adding particles selected from the modified Poisson distribution, Eq. (3).}}
\label{fig:figure3}
\end{figure}

At intermediate times, the interface width grows more slowly, with another growth exponent, $\beta_2$, which is not in the same universality class of the random deposition model. Figure \ref{fig:figure2} shows that this exponent depends on the lattice size L, specially for the smallest system sizes, which is a typical finite size behavior. For larger system sizes $\beta_2$ smoothly increases with L, reaching a limiting value when $L \rightarrow \infty$. In Figure \ref{fig:figure3} we show the plot of the growth exponent $\beta_2$ as a function of 1/L. The extrapolation to $L \rightarrow \infty$ gives the best value for this exponent, $\beta_2 = 0.310$. Fortuitously, it is close to the value 1/3, which is the exact result for the ballistic deposition in (1+1) dimensions, described by the KPZ equation.

On the other hand, at long times, when the surface width reaches its saturation value for each lattice size, we can estimate through equation \ref{family_vicsek} the value of the corresponding roughness exponent. We have found that $\alpha$ = 0.94$\pm$0.01. Another method to determine $\alpha$ is by using the rms fluctuation of the squared roughness in the steady state, presented in Refs. \cite{fabio_2, fabio_silveira}. This procedure gives an estimate for the roughness exponent, which is much less dependent on the finite-size corrections. The effective roughness exponent is defined as

\begin{equation}
 \alpha(L) \equiv \frac{1}{2} \frac{ln[\sigma(L)/\sigma(L/2)]}{ln2},
 \label{alpha}
\end{equation}
where $\sigma$ is defined by 

\begin{equation}
 \sigma \equiv \sqrt{<{w_{sat}}^2>-<w_{sat}>^2}.
\end{equation}

Using this method we have found that $\alpha$ = 0.942, which is in good agreement with our estimate based on the equation \ref{family_vicsek}. With the values of $\alpha$ and $\beta_2$, the dynamical critical exponent is $z=3.00$, which comes from the Family-Vicsek scaling relation. After deposition, particles can not migrate either downwards or upwards, and there is no surface diffusion in the present model. The exponents we calculated, $\alpha$, $\beta_2$ and $z$ fit very well to the nonlinear model, with conservative dynamics and non conservative noise in (1+1) dimensions, which is described by the Villain-Lai-Das Sarma equation \cite{vlds}. The exponents of this model were exactly determined from the renormalization group recurrence relations, where the nonlinear term determines the scaling behavior and the upper critical dimension is $d_c=4$. For $d=1$, the exponents are: $\alpha=1$, $\beta=1/3$ and $z=3$. In the deposition problems, we say that a given relaxation process is conservative if it does not change the number of particles in the system. On the other hand, a noise is said to be nonconservative when its correlation function at coordinates $(x,t)$ and $(x^{\prime}, t^{\prime})$ is of the white noise type, that is, it is the product of two Dirac's delta functions: one of them, is a function of the difference between the two spatial positions $(x-x^{\prime})$, and the other a function of the difference between the two times, $(t-t^{\prime})$. 

\section{Identical particles}

For surfaces formed by depositing identical particles, only particles with size in the range 2 to 9 lattice units of the linear substrate were considered. In our simulations, the particles are flat, with variable horizontal length (\textit{2 $\leq$ N $\leq$ 9 }) and vertical height corresponding to one unit. The deposition rules are exactly the same as the ones presented in the previous Section. The dependence of the surface width on the size of the particle is very weak and its behavior as a function of time for different lattice sizes is very similar to that seen for the deposition of particles of different sizes. We also find two different growth regimes characterized by the exponents $\beta_1$ and $\beta_2$. In Table I we display for each particle length ($N$), the values of the exponents $\beta_1$, $\beta_2$, $\alpha$ and $z$ for the lattice size $L=8192$. Then, even when the particles are of the same size, we find the this model is in the same universality class of the nonlinear model with conservative dynamics and non conservative noise in (1+1) dimensions.

\begin{table}[!htb]
 \begin{footnotesize}
  \begin{tabular}{c|c|c|c|c}
   \hline
   \hline
   \textit{Particle size (N)}& \multicolumn{4}{c}{\textit{L = 8192}}  \\ \cline{2-5}
            & $\beta_1$      &  $\beta_2$     & $\alpha$        & $z$\\ 
   \hline \hline
   2 	    & 0.50$\pm$0.04 & 0.32$\pm$0.02 & 0.95$\pm$0.01 & 2.98 \\
   3 	    & 0.50$\pm$0.02 & 0.32$\pm$0.02 & 0.93$\pm$0.01 & 2.94 \\
   4 	    & 0.51$\pm$0.03 & 0.32$\pm$0.01 & 0.94$\pm$0.01 & 2.94 \\
   5 	    & 0.51$\pm$0.07 & 0.32$\pm$0.03 & 0.94$\pm$0.01 & 2.92 \\
   6 	    & 0.50$\pm$0.03 & 0.31$\pm$0.04 & 0.95$\pm$0.02 & 3.05 \\
   7 	    & 0.50$\pm$0.01 & 0.31$\pm$0.04 & 0.97$\pm$0.01 & 3.07 \\
   8 	    & 0.51$\pm$0.04 & 0.32$\pm$0.03 & 0.95$\pm$0.01 & 2.96 \\
   9 	    & 0.52$\pm$0.03 & 0.33$\pm$0.02 & 0.95$\pm$0.01 & 2.89 \\
   \hline \hline
   \end{tabular}
  \end{footnotesize}
  \caption{\textit{The growth exponents for  different paticles' sizes.}}
 \label{tabela_N}
\end{table}

\section{Porosity}

There are few studies in the literature considering the formation of voids inside the volume \cite{karmakar, fabio_silveira}, and only deposits formed by the deposition of particles with two sizes have been analysed. In this section we present the results for the global porosity, where particles of different sizes are taken into account. The porosity, defined by Eq. \ref{porosidade} and calculated at each Monte Carlo step, saturates at the early stages of deposition, and exhibits a strong dependence only on the size of the particles. For deposits formed by a mixture of particles, selected from the modified Poisson distribution, or by identical particles with size in the range $ 2 \leq N \leq 9$, we didn't observe any dependence of the porosity on the size $L$ of the linear substrate, for lattice sizes larger than $L=512$.

\begin{figure}[t]
\vspace{0.7cm}
\includegraphics[scale=0.63]{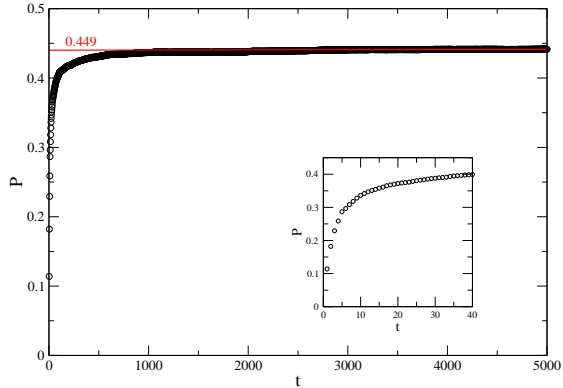}
\caption{\textit{Porosity versus time, for a bulk formed by depositing particles of different sizes on a linear lattice of size $L=4096$. The particles are chosen from the modified Poisson distribution, Eq. (3). The inset shows the porosity at the initial times.}}
\label{fig:figure4}
\end{figure}

Fig. \ref{fig:figure4} shows the time evolution of porosity for a deposit formed by a mixture of particles, whose sizes were selected from the modified Poisson distribution, Eq. (3), with $x=5$ and $n$ ranging from $1$ to $9$. The particles are aggregated on a linear substrate of size $L=4096$, and the resulting deposit is a two-dimensional structure. Just after \textit{50} time units we reach the stationary porosity value $P = 0.445 \pm 0.004$.

When we consider the formation of pores due to the deposition of identical particles with size larger than one lattice spacing, we find that the stationary porosity increases with the size of the particle. The deposition of particles with size exactly equal to one lattice spacing of the substrate does not lead to the formation of pores because the deposition follows the rules of the random deposition model. The inclusion of a very small fraction of particles of size \textit{2} is sufficient to generate a porous structure, which percolates over the whole deposit \cite{fabio_silveira}.

\begin{figure}[t]
 \centering
 \includegraphics[scale=0.31]{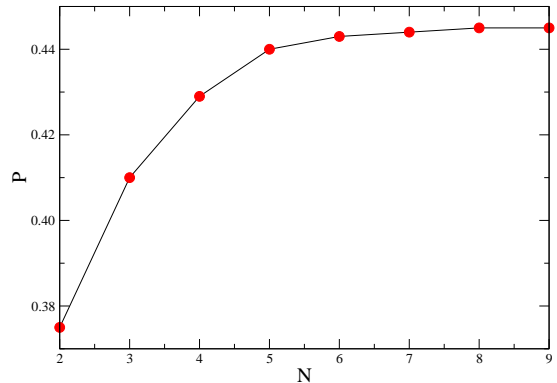}
 \caption{\textit{Porosity versus particle size N in a lattice with $L=4096$.}}
 \label{fig:figure5}
\end{figure}

As to be expected, the porosity increases with the size of the deposited particle. When a large particle is incorporated to the substrate, a shadow zone is created below it as we have seen in Fig. \ref{fig:figure1}. In this way, we expect the porosity increases with the size of the particle. We show in Fig. \ref{fig:figure5} the behavior of the stationary porosity as a function of the size of the deposited particle.

We note that the increase of the porosity is large for small particle sizes, but it reaches a value near its saturation for particles of intermediate size. It is easy to understand that the saturation value for this problem is $P=0.5$. A simple reasoning shows that the maximum value of $P$ is 0.5. This happens when the size of the particle to be deposited is $N=L/2$, i.e., half of the lattice size. After a particle of this size is incorporated to one of the ends of the substrate, the probability that the next particle is deposited at the same height as the former one is about $1/L$, which is a very small number when we take larger substrates. On the other hand, if the particle is not deposited exactly at one of the ends of the linear substrate, no other particle can be incorporated with the same height. Then, in this case, only a single particle is deposited at each Monte Carlo step and we have $P=1/2$. For values of $L/2 < N < L$, the porosity decreases, and is given by 

\begin{equation}
 P = \frac{L - N}{L},
 \label{porosidade_maxima}
\end{equation}
and only a single particle is aggregated in each Monte Carlo step, which happens, in fact, in the first trail of deposition. Our results show that, even for particles of small size, as for instance $N=9$, we find that $P=0.489$ for $L=8192$, which is very close to the maximum possible value, $P=0.5$.

\section{Conclusions}

We studied a surface growth model where particles of different sizes are deposited on a linear substrate. Particles with sizes in the range $1 \leq N \leq 9$ are selected from a modified Poisson distribution and fall from random positions over an initially flat surface. We also considered the deposition of identical particles with size larger than one lattice spacing of the linear substrate. Through Monte Carlo simulations we found that the surface width evolves in time following there different behaviors. At the initial times the surface roughness behaves as in the random deposition model, and the growth exponent is
$\beta_1 = 0.50$. At intermediate times, the surface roughness grows more slowly with the exponent $\beta_2 = 0.310$. Finally, at long times, it enters into the saturation regime, characterized by the roughness exponent, $\alpha = 0.94$. The estimate dynamic exponent is $z=3.0$. These figures put the present deposition model in the same universality class of the nonlinear model with conservative dynamics and non conservative noise in $(1+1)$ dimensions.
The two-dimensional bulk formed by the deposition of particles larger than one cell unit of the linear substrate reveals us a porosity increasing very fast at the initial times, and also reaching a saturation value, which depends on the size of the particle. For particles of size $N=9$ the stationary porosity reaches a value which is $2\%$ away of the maximum possible value $P=0.5$.


\section*{Acknowledgements}
The authors would like to thank Conselho Nacional de Desenvolvimento Cient\'ifico e Tecnol\'ogico (CNPq) by the scholarships.


\end{document}